# Direct observation of multi-spinon excitations outside of the two-spinon continuum in the antiferromagnetic spin chain cuprate $Sr_2CuO_3$

J. Schlappa[1,2]*, U. Kumar[3], K. J. Zhou[2,4], S. Singh[5], M. Mourigal[7]‡, V. N. Strocov[2], A. Revcolevschi[6], L. Patthey[2], H. M. Rønnow[7], S. Johnston[3]*, and T. Schmitt[2]*

[1]European X-Ray Free-Electron Laser Facility GmbH, Schenefeld, Germany.
[2]Swiss Light Source, Paul Sherrer Institut, CH-5232 Villigen PSI, Switzerland.
[3]Department of Physics and Astronomy, The University of Tennessee, Knoxville, Tennessee 37996, USA.
[4]Diamond Light Source, Harwell Science and Innovation Campus, Didcot, Oxfordshire.
[5]Indian Institute of Science Education and Research, Pashan, Pune, India.
[6]Université Paris-Sud, Orsay Cedex, France.
[7]École Polytechnique Fédérale de Lausanne (EPFL), Lausanne, Switzerland.
‡Present address: School of Physics, Georgia Institute of Technology, Atlanta, GA 30332, USA.

*Corresponding authors: justine.schlappa@xfel.eu, sjohn145@utk.edu, thorsten.schmitt@psi.ch

**One-dimensional (1D) magnetic insulators have attracted significant interest as a platform for studying emergent phenomena such as quasiparticle fractionalization and quantum criticality. The antiferromagnetic Heisenberg chain of spins-1/2 is an important reference system; its elementary excitations are spin-1/2 quasiparticles called spinons that are always created in pairs. However, while inelastic neutron scattering (INS) experiments routinely observe the excitation continuum associated with two-spinon states, the presence of more complex dynamics associated with four-spinon states has only been inferred from comparison with theory. Here, we show that resonant inelastic x-ray scattering (RIXS) is capable of accessing the four-spinon excitations directly, in a spectroscopic region separated from the two-spinon continuum. Our results provide the first direct measurement of four-spinon excitations, which is made possible by the fundamentally different correlation functions probed by RIXS compared to INS. This advance holds great promise as a tool in the search for novel quantum states, in particular quantum spin liquids.**

**Main Text**

**Introduction -** When confined to one spatial dimension (1D), systems of interacting electrons host an assortment of macroscopic many-body phenomena, including quantum critical magnetic states with collective excitations carrying fractional quantum numbers. For this reason, quasi-1D magnetic insulators have attracted wide experimental and theoretical interest as an ideal playground for studying quantum many-body phenomena. Owing to numerous experimental realizations of quantum states in real materials, some of the most stringent tests of quantum many-body theory have been conducted in 1D [1-17].

The 1D Heisenberg antiferromagnet (HAFM), where localized spins $S$ interact with their nearest-neighbours via an exchange interaction $J$, is perhaps the simplest of these systems; the spin-1/2 case is an important reference system that can be solved exactly using the Bethe ansatz. The ground state is a macroscopic SU(2)-symmetric singlet, in which quantum fluctuations suppress long-range order, leading to a spin liquid ground state even in the limit of zero temperature. The elementary excitations are collective spin fluctuations called spinons, which carry spin ½ but no charge. Spinons are created in pairs by elementary spin-flip processes generated during inelastic neutron scattering (INS) or resonant inelastic x-ray scattering (RIXS) experiments. A continuum composed of two-spinon states dominates the excitation spectrum of the HAFM, as directly observed on different realizations of the 1D HAFM by INS [2,3,5] and RIXS [1,18,19]. However, two-spinon excitations do not provide a complete basis to describe the dynamics and weaker continua of four-, six-, …-spinon states are also predicted and understood through applications of the Bethe ansatz [1,5] or density matrix renormalization group [10,11].

The observation of such higher-order spinon continua is an ongoing area of research [2,14,15,19]. To date, the contribution of the four-spinon excitations to the magnetic dynamic structure factor $S(q,\omega)$ of the 1D HAFM has been *inferred* from a comparison between the measured and calculated spectral weight of the two-spinon continuum. For example, Ref. [2] measured the magnetic excitation spectrum of the 1D HAFM $CuSO_4 \bullet 5D_2O$ with INS in absolute units and found that the majority of the spectral weight was contained within the boundaries of the two-spinon continuum. (Similar results have been obtained for other 1D HAFM systems [10].) The calculated $S(q,\omega)$ using the Bethe ansatz reveals that the two-spinon excitations account for 74% of the total detected weight, with the remaining 26% produced predominantly by four-spinon contributions. Moreover, the four-spinon contributions in these studies were solely found in regions of momentum space overlapping with the two-spinon states, while the allowed phase space for four-spinon excitations is in fact much larger [19]. Indeed, four-spinon excitations have also been (controversially) reported outside of the two-spinon continuum in an INS study of a frustrated 1D ferromagnet $LiCuVO_4$ near $q = \pi/2a$ [14-16]. This result raises the question whether higher-order spinon excitations can be observed directly in the pure 1D HAFM, namely in regions of phase space separated from the two-spinon continuum? Here, we show that RIXS at the O *K*-edge allows for such observation, a capability that results from the fundamentally different correlation function it probes compared to other techniques.

RIXS is a photon-in photon-out spectroscopy technique where photons inelastically scatter from a sample [8]. In a RIXS experiment, the photon energy $\hbar\omega_{\text{in}}$ of the incident X-rays is tuned close to an absorption edge of an atomic species in the material of interest, thereby initiating an electron transition between a core level and an unoccupied valence band state. This process creates an intermediate state with an additional electron either in the valence or conduction band and a hole in the core level. This core-hole excited state will decay on a femtosecond time scale, leaving the system in a valence-band excited state. Since X-ray photons carry substantial momentum (in contrast to the light of optical or VUV wavelengths), these excited states can be studied both in the energy and the momentum domain. Thus, RIXS can be viewed as momentum-resolved resonant Raman spectroscopy, suitable for mapping dispersions of excitations in quantum materials.

In recent years, RIXS has been used to probe electronic excitations involving charge [1,17,20], orbital [1,17], spin [1,12,20-26], and lattice [13,26] degrees of freedom in a wide range of materials. Studies on the dynamic magnetism have largely focused on cuprates, where direct spin-flip excitations can be investigated at the Cu $L_3$-edge due to strong spin-orbit coupling in the core levels [27]. Indeed, in Cu $L_3$ RIXS measurements of the quasi-1D spin chain cuprate $Sr_2CuO_3$, two-spinon continuum excitations could be probed (with indications of also four-spinon excitations) [1]. Studies in other cuprate materials revealed two-triplon excitations in the spin-ladder system $Sr_{14}Cu_{24}O_{41}$ [12] and magnon excitations in many quasi two-dimensional superconducting cuprates [21,22,26,27]. In this article, we report momentum-resolved oxygen $K$-edge RIXS studies of the quasi-1D spin-chain cuprate $Sr_2CuO_3$, one of the best realizations of the 1D HAFM (Fig. 1**a**). We observe magnetic excitations that exist in two non-overlapping regions of phase space. Through detailed modeling within the $t-J$ model, we show that these two distinct sets of excitations correspond to the two- and four-spinon continua. Specifically, four-spinon excitations centered at 500 meV energy loss give a strong and broad response around the Γ-point ($q=0$) that is separated from the more commonly observed two-spinon continuum. Our results establish a new channel for the creation of magnetic excitations in 1D materials, not present in other probes (e.g. in INS). This capability stems from the dynamics of the intermediate state, which grants access to fundamentally different correlation functions.

**Results –** The low-energy electronic degrees of freedom in the charge transfer insulator $Sr_2CuO_3$ are formed from the $CuO_4$ plaquettes, which are arranged into 1D corner-shared chains [24], as shown in Fig. 1**a**. In the atomic limit, the Cu ion is in a *d⁹* valence state, with a single hole occupying the Cu $3d_{x^2-y^2}$ orbital. There is, however, significant hybridization between the Cu *3d* and *2p* orbitals of the surrounding oxygen, resulting in a substantial superexchange interaction $J \sim 250$ meV [1,4,5] between the Cu spins. In the real material, the individual -Cu-O-Cu- chains are weakly coupled such that the system has a bulk Neel temperature of $T_N \sim 5$ K [24]. Above this temperature, however, the chains decouple and become nearly ideal realizations of the 1D Heisenberg AFM, as evidenced by the observation of the two-spinon continuum in INS [5] and Cu $L_3$ RIXS [1]. The same RIXS study [1] also found evidence for novel spinon-orbiton separation effects in $Sr_2CuO_3$, further underscoring the importance of the 1D physics.

Figure 1**b** shows the x-ray absorption (XAS) data of $Sr_2CuO_3$ measured at the O K edge (a $1s \rightarrow 2p$ resonance). The intensity reflects the partial density of the unoccupied valence and conduction band states, here projected onto the oxygen orbitals. We observe a sharp excitonic structure in the pre-edge region and broad continuum states at energies above 529 eV. The excitonic peak corresponds to excitations of the O 1s core electron into the upper Hubbard band (UHB), creating a Cu $3d^{10}$ state [28]. This excitation is allowed by the sizable hybridization between the O *2p* and Cu *3d* orbitals. The UHB XAS peak depends strongly on the polarization of the incident photons, reflecting the strong anisotropy of the system [28]. In particular, the suppression of intensity for σ-polarized light indicates that the unoccupied states are oriented in the plane of the $CuO_4$ plaquettes, whereas the energy shift upon changing the incidence angle to π-polarized light reflects differences in coordination between the out-of-chain and the in-chain oxygens (indicated in Fig. 1**a** as sites A and B, respectively), in agreement with previous findings [29]. For the remainder of this work, we focus on RIXS spectra recorded with the incident photon energies tuned to the UHB B (or A) peak, where an in-chain (or out-of-chain) O *1s* core electron is promoted into a neighbouring Cu *3d* orbital. This final state of the XAS process dictates the intermediate state of RIXS, and is important in determining the scattering cross-section.

Figure 1**c** shows RIXS spectra measured with the photon energy tuned to the resonance of the A and B peak at the O *K*-edge in comparison with Cu $L_3$-edge data at $q = \pi/2a$. There are two energy regions with pronounced excitations: one below 1 eV and one above 1.5 eV, separated by a region of very weak spectral weight. The excitations at higher energies are dominated by *dd* and charge transfer excitations (CT)

[1,4]; the *dd* excitations are dominant at the Cu $L_3$-edge, whereas the charge transfer excitations are dominant at the O *K*-edge. Figure 1**d** zooms in on the low-energy excitation region, which is our focus. O *K* RIXS for photon energies tuned to B with different incident angles are compared to low-energy Cu $L_3$ RIXS data.

Below 1 eV we see several excitations. In addition to the elastic line at zero energy loss, we also observe an excitation at ~ 90 meV with varying cross section for the different configurations, which corresponds to a bond-stretching phonon [13]. A line spectrum at $q = \pi/2a$ (Fig. 2**d**) reveals a sharp structure coinciding with the very strong two-spinon excitations at the same $q$-point in Cu $L_3$-edge data. In addition, a line spectrum taken close to the Γ-point (Fig. 2**e**) is dominated by a broad structure, centered at ~0.5 eV and extending to about 1 eV in energy loss. The energy of this structure is well separated from the *dd*- and CT excitations, suggesting that they are magnetic in origin. To probe the character of these low-energy magnetic excitations in the O *K*-edge RIXS spectra, we have studied their momentum dependence, as shown in Fig. 2**a**. (O *K*-edge RIXS allows studying about 25% of the first BZ along [100] towards each side of $q = 0$, see Fig. S1.) In addition to the strong phonon excitation in O *K*-edge data, there are two distinct sets of continua in the magnetic region between 0.2 eV and 1.0 eV. One is dispersing towards zero energy for $q = 0$ and lies well within the boundaries of the two-spinon continuum (indicated by the white dotted lines). The second region is centered at $q = 0$ and 500 meV energy loss, and is situated clearly outside of the two-spinon continuum. Comparison to Cu $L_3$ data displayed in Fig. 2**c**, where the two-spinon continuum dominates the spectrum, illustrates that O *K*-edge and Cu $L_3$-edge RIXS have quite different response to magnetic excitations. In addition, the O K-edge data reveal much stronger polarization dependence due to difference in connectivity of the in-chain and out-of-chain O 2p orbitals (see Supplementary Note 1 and Fig. S1). However, the line cuts of O K-edge and Cu $L_3$-edge RIXS spectra at $q = 0$ in Fig. 2**f** show that there is also a finite weight in Cu $L_3$-edge RIXS spectra.

We performed small cluster exact diagonalization (ED) calculations to elucidate the nature of the excitations at the Γ-point. Since we are interested in the energy region well below the *dd*- and CT-excitations, we used the $t - J$ model, where these processes have been integrated out. The computed spectra (with elastic peak removed) are compared against the experimental data in Fig. 2**b**. Line cuts of the data superimposed over the calculations are shown in Figs. 2**d** ($q \approx \pi/2a$) and 2**e** ($q = 0$). The overall agreement between the calculations and the experimental data is excellent. (Note that the phonon excitation is not included in the theory.) The level of agreement indicates that the final states of the O K-edge RIXS process can be well described solely by excitations of the half-filled $t - J$ model, whose final states are the same as those in the Heisenberg model. We can therefore identify the upward dispersing branch as the two-spinon continuum (with some small contribution from four-spinon excitations) while the continuum of excitations centered at $q = 0$ corresponds to four-spinon (FS) excitations. This assignment is further supported by the dependence of these excitations on the core-hole lifetime, which will be discussed shortly. Importantly, these FS excitations are observed directly and well separated from the two-spinon continuum.

**Discussion –** How can we understand the magnetic excitations in RIXS captured by the $t - J$ model, and why do we see contributions that are absent in INS? In Fig. 3 we illustrate the magnetic excitation mechanisms in a spin chain with the different scattering techniques: INS (Fig. 3**a**), Cu $L_{2,3}$- (Fig. 3**b**), and O *K*-edge RIXS (Fig. 3**c** & 3**d**). For INS, the total spin of the spin chain and the scattered neutron must be conserved. Flipping the spin of the scattered neutron must therefore be accompanied by a spin flip with $\Delta S = 1$ in the spin chain. In a simplified picture, such an excitation leads to the creation of two domain walls in the spin chain (Fig. 3**a**) that decay into two spinons carrying parallel spins.

Unlike INS (and RIXS at the Cu $L_{2,3}$ edges), single spin flip excitations with $\Delta S = 1$ are generally forbidden for *K*-edge RIXS[1]. Instead, $\Delta S = 0$ processes like the one sketched in Fig. 1**e** must be used to create magnetic excitations. Here, the incident photon creates a Cu $3d^{10}$ UHB excitation in the intermediate state, resulting in a Cu site with an additional "spin-down" electron in direct vicinity to an O *1s* core hole. The 180° Cu-O-Cu bonding angle in $Sr_2CuO_3$ enables efficient double inter-site hopping of *3d* electrons between two adjacent Cu sites via the bridging in-chain oxygen site (B in Fig. 1**a**), transferring the Cu $3d^{10}$ to the neighbouring Cu site. Since this Cu atom is also hybridized with the oxygen where the core hole is localized, the "spin-down" electron can then decay and fill the core level, leaving the system with a net double inter-site spin flip. This process, sketched in Fig. 3**c**, is analogous to an indirect double spin flip process predicted for Cu *K* RIXS [30]. In two dimensions, this process can only create bi-magnon excitations. In one dimension, it gives rise to a double domain wall that decays into two-spinons carrying antiparallel spins [19]. This excitation pathway explains the presence of the two-spinon continuum in O *K*-edge RIXS spectra. But how

---

[1] This statement holds only for materials with small spin-orbit coupling in the valence band; single-flips are allowed in O K-edge RIXS on iridates, see ref. 20)

can we visualize the scattering process responsible for creation of four-spinon excitations around the Γ-point? Here, the lifetime of the intermediate state plays a critical role.

In RIXS there is an intermediate core-hole excited state with a doublon in the Cu 3d shell. During the RIXS scattering process the angular momentum of the photon must be conserved. The selection rule for the allowed electronic states in the RIXS scattering processes depends among others on the strength of the spin-orbit coupling in the core state. In case of strong spin-orbit coupling, the total angular momentum ($J$) of an electronic state must be conserved rather than the spin. For a Cu 2p core-hole, the change of spin momentum can be compensated by the change of angular momentum, allowing for $\Delta S = 1$ spin-flip excitations in Cu $L_{2,3}$-edge RIXS (Fig. 3**b**) that are similar to INS [27]. In contrast to INS, however, RIXS involves a doublon in the intermediate state, which decays on a timescale set by the corehole lifetime (~ several fs) [31]. During this time, the additional charge in the intermediate state can interact with the system, creating excitations that are not possible in INS. For the O 1s core hole there is no appreciable angular momentum available; therefore, the spin momentum must be conserved and only $\Delta S = 0$ excitations are possible (as described above) (Fig. 3**c**). In a 1D system, the result of this $\Delta S = 0$ excitation looks very similar to the result of a single spin flip $\Delta S = 1$ in that both excitations lead to the creation of two domain walls, but at the O K-edge they are separated by at least one atomic site and have opposite spins. The lifetime of O 1s core-hole states is somewhat longer than the lifetime of Cu 2p core-hole states, however. During this time, the doublon in the 3d band can also generate double spin flips on the surrounding sites, as sketched in (Fig. 3**d**), creating two additional double spin flips separated by larger lattice distances. The subsequent decay of the core hole results in the creation of two additional domain walls, creating a total of four spinons in the final state. This scattering channel is the direct result of the additional charge in the intermediate state. Moreover, its intensity depends on the lifetime of the core-hole as a longer-lived doublon will have sufficient time to generate the longer range double spin-flips. This new excitation channel is expected to be weak in Cu $L_3$ RIXS, whose core-hole is short lived, and completely absent in INS.

We performed calculations for the dependence of these excitations on the lifetime of the intermediate state to test our interpretation. The results are presented in Fig. 4. We observe that upon decreasing the core-hole lifetime (increasing $\Gamma$) the intensity of magnetic excitations in O $K$-edge RIXS decreases. Moreover, the spectral weight of the FS continuum moves towards smaller energy losses (see Fig. 4**a**). The decrease in intensity is much slower for excitations belonging to the two spinon continuum than for the FS excitations. Whereas the two-spinon excitations are still well pronounced for $\Gamma = 500$ meV (b and c), the FS excitations are suppressed below $\Gamma = 300$ meV (a), which is comparable to the exchange interaction $J$. The suppression of $q = 0$ FS weight proves that the core-hole lifetime sets the time scale for the intermediate state to generate FS excitations. As its lifetime is quenched below $J$ (~ 1.3 fs), there is not enough time for additional double spin-flips to occur in the chain during the frustrating presence of the doublon. The dynamics of this intermediate state plays therefore an important role for the discovered excitation channel for magnetic excitations and produces additional selection rules – beyond a single or double spin-flip.

**Conclusions –** We have demonstrated that RIXS produces complementary selection rules for magnetic scattering to INS, which arises from the lifetime and dynamics of the intermediate state. Importantly, the new selection rule is unique to RIXS and provides access to non-local spin correlation functions beyond two-site correlation functions probed by traditional scattering techniques. O *K*-edge RIXS has long core-hole lifetimes and is therefore ideal for examining excitations "beyond" INS scattering, as long lifetimes of the intermediate state allow charge fluctuations to take place. We have exploited this fact to observe *directly* the four-spinon excitations of a 1D AFM, located outside the two-spinon continuum for the first time. This technique opens a completely new avenue to explore quantum magnetism and quasi-particle fractionalization, which has broad applications in the field of quantum magnetism. Time-resolved studies at the upcoming x-ray free-electron laser (XFEL) sources, e.g. European XFEL and Swiss FEL, will hopefully facilitate studying such dynamics at the fs-timescale.

**Methods**

**Experiment –** We applied the technique of high-resolution resonant inelastic x-ray scattering (RIXS) with the incident photon energy tuned to the O 1s core → 2p UHB resonance (around 528 eV). Single-crystal samples of $Sr_2CuO_3$ were grown by the floating-zone method and freshly cleaved before the RIXS experiment. During the experiment the surface normal to the sample, [010], and the propagation direction of the chains, [100], were oriented parallel to the scattering plane. The scattering plane was horizontal. The sample was cooled with a helium-flow cryostat to 14 K during the measurements. The experiments were performed at the ADRESS beamline of the Swiss Light Source at the Paul Scherrer Institut [32,33]. Incident photons were linearly polarized either in the scattering plane ($\pi$-polarization), which was the case for most of the data, or perpendicular to the scattering plane ($\sigma$-polarization). The XAS data was measured in total fluorescence yield. The beamline (BL) energy resolution was set to 70 meV or better, with the BL exit slit open to 30 µm (the BL energy resolution for the Cu $L_3$ data [1] was 100 meV or better, with the BL exit slit open to 10 µm.) The RIXS spectrometer was located at a fixed scattering angle of $\Psi$ = 130° ± 1°, whereas the incidence angle on the sample varied between 10° ± 1° and 110° ± 1° grazing (see Fig. S1). The angular horizontal acceptance of the spectrometer was approximately 5 mrad [33]. The total experimental energy resolution was 80 meV and the simultaneously recorded energy window was 22.2 eV (the total experimental resolution for the Cu $L_3$ data [1] was 140 meV and the simultaneously recorded energy window was 59.2 eV).

**Cluster Calculations** – The RIXS intensity $I(q, \Omega)$ was evaluated using the Kramers-Heisenberg formalism where ($\hbar = 1$)

$$I(q,\Omega) = \sum_f \left| \sum_{n,R_m} e^{-iqR_m} \frac{\langle f|D_m^\dagger|n\rangle\langle n|D_m|i\rangle}{E_i + \omega_\text{in} - E_n + i\Gamma} \right| \delta(E_f - E_i + \Omega)$$

Here, $q = k_\text{out} - k_\text{in}$ is momentum transfer and $\Omega = \omega_\text{out} - \omega_\text{in}$ is the energy loss, $D$ is the dipole operator, and $|i\rangle$, $|n\rangle$, and $|f\rangle$ are the initial, intermediate, and final states of the RIXS process with energies $E_i$, $E_n$, and $E_f$, respectively, $R_m = am$ is the position of the $m^\text{th}$ Cu atom, $a$ is the Cu-Cu distance, and $\Gamma$ is the core-hole lifetime. We compute the eigenstates by diagonalizing $t - J$ Hamiltonian defined on a twenty-two site cluster.

The use of this low-energy effective model is justified since all of the *dd* and charge-transfer excitations appear well above 1 eV in energy loss (see Fig. 1c). The dipole operator is given by

$$D_m = \sum_\sigma (d_{m,\sigma} - d_{m+1,\sigma}) s^\dagger_{m,\sigma}$$

where $d_{m,\sigma}$ annihilates a spin $\sigma$ hole on Cu site $m$ and $s^\dagger_{m,\sigma}$ creates a hole in the oxygen 1*s* orbital on the site between the $m$ and $m+1$ Cu sites. Here, the relative phases reflect the phases of the original Cu-O overlap integrals. The model parameters are $t$ = 300 meV and $J$ = 250 meV, which is appropriate for Sr$_2$CuO$_3$ [4], and $\Gamma$ = 150 meV for the oxygen K edge [13,17].

**Acknowledgements:** The authors thank C. Batista, and K. Wohlfeld for useful discussions. The experiments were performed at the ADRESS beamline of the Swiss Light Source at the Paul Scherrer Institut. We acknowledge support from the Swiss National Science Foundation and its NCCR MaNEP. CPU time was provided in part by resources supported by the University of Tennessee and Oak Ridge National Laboratory Joint Institute for Computational Sciences (http://www.jics.utk.edu).

**Author Contributions:** J. S. and T. S. designed the experiment. J. S., T. S., and K. J. Z. performed the experiment with the assistance of V. N. Strocov. J. S. performed data analysis in discussion with T. S., U. K. and S. J. performed theory calculations. S. S. and A. R. have grown the samples. T. S. and S. J. were responsible for project management. J. S. wrote the paper with input from all authors.

**Competing Interests:** The authors declare that they have no competing financial interests.

**Correspondence**: Correspondence and requests for materials should be addressed to S. J. (email: sjohn145@utk.edu), J. S. (email: justine.schlappa@xfel.eu), or T. S. (email: thorsten.schmitt@psi.ch).

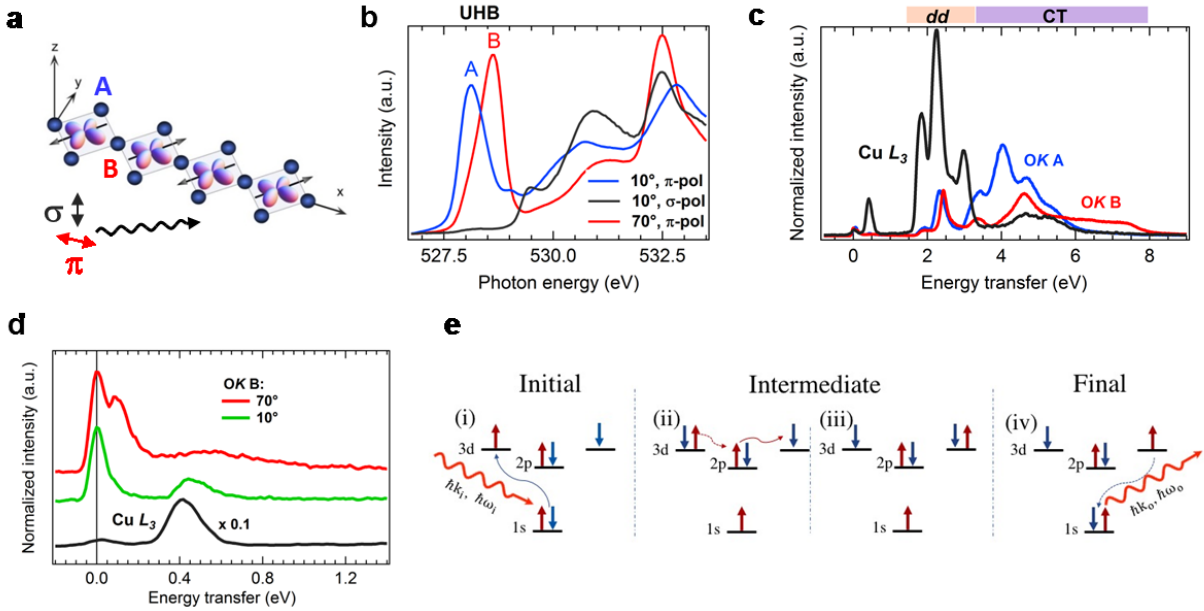

**Figure 1 – Summary of the experimental data at the oxygen *K*-edge. a** Cartoon sketch of the Cu-O-Cu corner-shared chains forming the active low-energy degrees of freedom in $Sr_2CuO_3$, and of the incident-light geometry. The Cu atoms are primarily in a $d^9$ valence state, where a single hole occupies each of the Cu $3d_{x^2-y^2}$ orbitals and interacts antiferromagnetically with its in-chain neighbors. **b** The polarization dependence of the XAS spectra. $\sigma$-polarized light probes unoccupied states perpendicular to the $CuO_4$-plaquettes, having no spectral weight at the UHB (there are no apical oxygens). Data obtained with $\pi$-polarized light at incidence angles of 70° (close to normal incidence and $q \approx 0$) and 10° (grazing incidence geometry and $q \approx \pi/2a$) primarily probes the out-of-chain (A) and in-chain (B) oxygen sites, respectively. The difference in the pre-peak resonance corresponds to the differences in the chemical environments of these two oxygen sites (chemical shifts), where the B site hosts the plaquette-connecting oxygen orbital [28]. **c** Polarization dependence for $\pi$-polarized O *K*-edge RIXS data for incident energies tuned to the A and B peaks in the XAS shown in panel **b** (incident angles as in **b**). The $\pi$-polarized Cu $L_3$-edge RIXS data at 20° incidence angle ($q \approx \pi/2a$) is also shown for comparison. The RIXS spectra are normalized to acquisition time. The peaks above 1.8 eV are associated to *dd* (orbiton) and charge transfer (CT) excitations, as indicated. The peak below 0.6 eV in the Cu $L_3$ data corresponds to multi-spinon excitations [1]. **d** The Cu $L_3$ and O *K* B-resonance RIXS data from panel **c** plus B-resonance for 10° incidence angle ($q \approx \pi/2a$), now focusing on the first 1.3 eV energy loss, where several low-energy spin excitations are found. **e** Sketch of the double spin-flip process across two Cu sites at the oxygen *K*-edge.

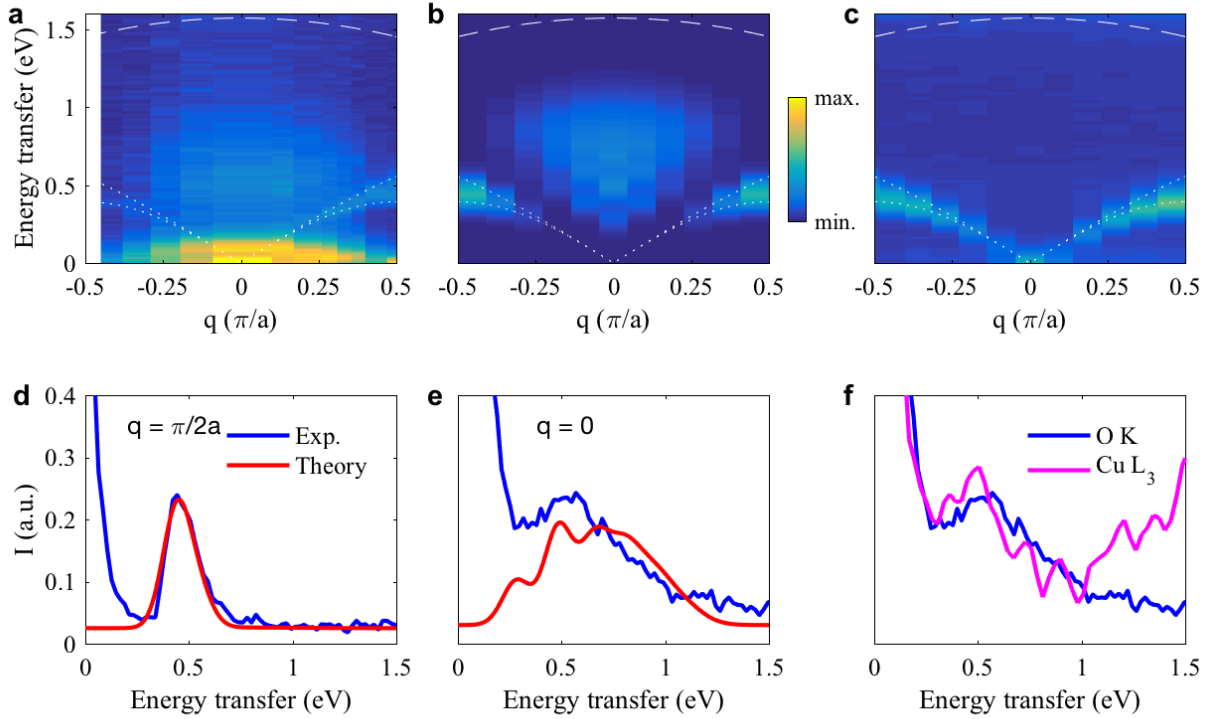

**Figure 2 – Comparison between the experimental and calculated RIXS spectra at the oxygen *K*-edge. a** The measured and **b** calculated RIXS spectra at the oxygen – edge as a function of momentum transfer and energy loss. The measured spectra were obtained with an incident photon energy of $\hbar\omega_{in} = 528.6$ eV (resonance B) while the calculated spectra is for $\hbar\omega_{in} = 500$ meV. (This value optimizes the intensity of the four-spinon features, see Supplement.) Panel **c** shows the measured RIXS spectra at the Cu $L_3$-edge after Ref. [1]. The dotted and dashed white lines in panels a-c indicate the boundaries of the two- and four-spinon continua, respectively. The excitation at ~90 meV in the Oxygen *K*-edge data is a phonon excitation not included in our model calculations. The modeled RIXS intensity was obtained from exactly diagonalizing a 22-site $t-J$ chain with periodic boundary conditions and the elastic line has been removed from the data for clarity. Panels **d** and **e** show line cuts of the RIXS spectra at $q = \pi/2a$ and $q = 0$, respectively. Panel **f** compares the O *K*-edge and Cu $L_3$-edge RIXS spectra at $q = 0$. In the case of Cu $L_3$ data, there is a tailing contribution from higher energy *dd* excitations, which extends down to low energy loss (see Supplement).

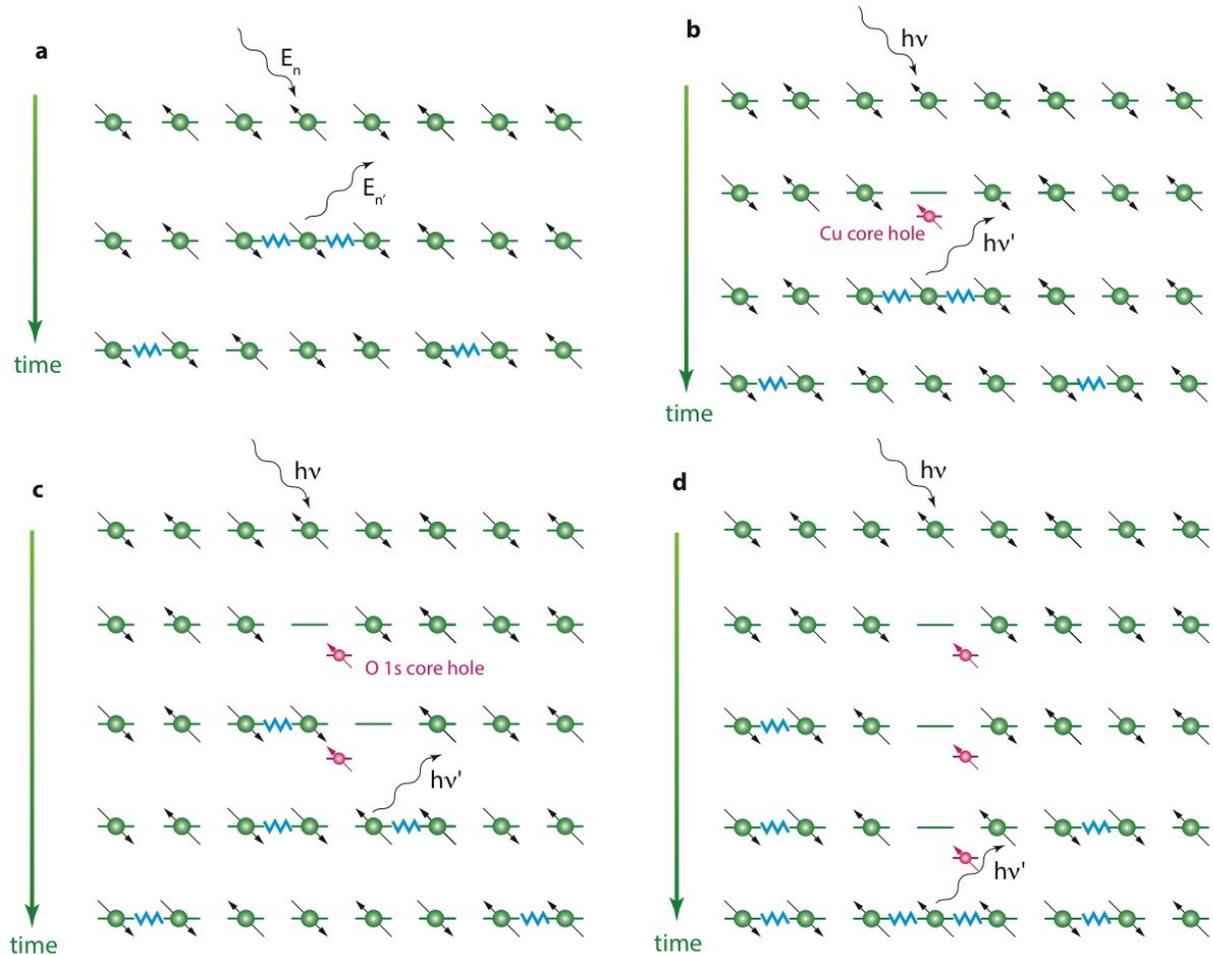

**Figure 3 – An illustration of various spin excitation processes. a** The $\Delta S = 1$ direct spin flip process that occurs in an inelastic neutron scattering experiment, which primarily decays into two-spinon excitations that are visualized as domain walls in the AFM background [2]. **b** The same $\Delta S = 1$ spin flip process in RIXS, which is accessible in materials with strong spin-orbit coupling in the core level [27]. **c** The indirect double spin-flip process at the oxygen *K*-edge, which occurs via the multi-orbital hopping processes sketched in Fig. 1**e**. This process generates a nearest-neighbour double spin flip, which predominantly decays into a two-spinon excitation [30]. **d** A second order process at the oxygen K-edge that produces four-spinon excitations. Here, the absence of the spin in the AFM chain allows double spin flips to occur on the sites adjacent to the missing spin. These double spin flips generate spinon excitations away from the site where the core hole is created. The subsequent decay of the core hole then produces two additional spinons in its vicinity. This process requires a long-lived core-hole to allow for sufficient time to generate the two double spin-flips before the core-hole decay occurs.

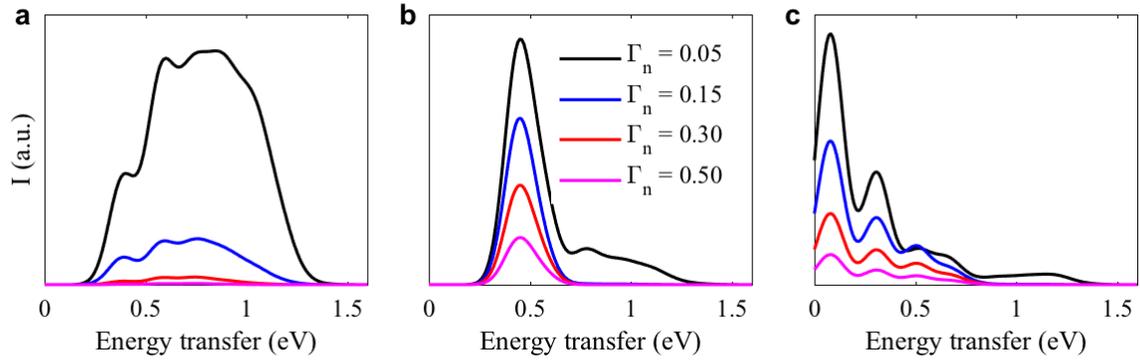

**Figure 4 – The effect of the core-hole lifetime on the RIXS spectra.** The variation in the computed RIXS intensity at **a** $q = 0$, **b** $q = \pi/2a$, and **c** $q = \pi/a$. As the core-hole lifetime is decreased (increasing $\Gamma$), the four-spinon excitations at $q = 0$ disappear rapidly, while the two-spinon contributions to the spectra at $q = \pi/2a$ and $q = \pi/a$ are more robust.